\documentclass[conference]{IEEEtran}
\usepackage{eso-pic}
\usepackage{xcolor}
\usepackage{placeins}
\usepackage{float}
\IEEEoverridecommandlockouts
\usepackage{cite}
\usepackage{amsmath,amssymb}
\usepackage{graphicx}
\usepackage{booktabs}
\usepackage{tabularx}
\usepackage{array}
\usepackage{makecell}
\usepackage{algorithm}
\usepackage[noend]{algpseudocode}
\usepackage{url}
\usepackage{balance}
\definecolor{acceptblue}{RGB}{0,51,102}

\newcolumntype{C}[1]{>{\centering\arraybackslash}p{#1}}

\newcommand{\yes}{\ensuremath{\checkmark}}
\newcommand{\no}{\ensuremath{\times}}
\newcommand{\partly}{\ensuremath{\triangle}}

\usepackage[hidelinks]{hyperref}
\newcolumntype{Y}{>{\raggedright\arraybackslash}X}
\newcolumntype{L}[1]{>{\raggedright\arraybackslash}p{#1}}
\algrenewcommand\algorithmicindent{0.8em}
\makeatletter\renewcommand{\ALG@beginalgorithmic}{\footnotesize}\makeatother
\begin{document}

\AddToShipoutPictureFG*{%
  \AtPageUpperLeft{%
    \raisebox{-12mm}{%
      \makebox[\paperwidth][c]{%
        \normalfont\footnotesize
\textcolor{acceptblue}{%
  This paper was accepted to IEEE IEMCON 2026%
}
      }%
    }%
  }%
}
\title{A2A-CaseVerify: Merkle-Linked Case-Evidence Verification for Cross-Organization A2A Workflows}

 \author{
\IEEEauthorblockN{
Adil Alshammari\textsuperscript{1,2}
\quad
Hayretdin Bahsi\textsuperscript{1,3}
}

\IEEEauthorblockA{
\textsuperscript{1}School of Informatics, Computing, and Cyber Systems,
Northern Arizona University, Flagstaff, Arizona, USA\\
\textsuperscript{2}College of Computer and Information Sciences,
Majmaah University, Majmaah 11952, Saudi Arabia\\
\textsuperscript{3}School of Information Technologies,
Tallinn University of Technology, Tallinn, Estonia\\
aha388@nau.edu;
hayretdin.bahsi@nau.edu
}
 }

\maketitle

\begin{abstract}
Agent-to-Agent (A2A) communication enables large language model (LLM) agents to exchange tasks, messages, and artifacts across organizations. A valid message alone does not establish that a final workflow claim is supported by a complete, ordered, and case-consistent evidence path. We present A2A-CaseVerify, a deterministic offline verifier that maps preserved A2A runtime objects, business events, and typed support edges to a directed case evidence graph. It commits A2A envelopes, event projections, and support edges, then combines event and edge leaves into a deterministic case Merkle root. The verifier returns \textsc{SUPPORTED} with status \texttt{OK} only when protection and profile-specific reconstruction checks pass. We evaluate one canonical supported healthcare-profile bundle and 13 controlled mutations. We separately test a valid bundle generated with the official Python A2A software development kit (SDK). In these tests, the canonical and SDK-generated bundles are accepted. All 13 mutations are rejected, and each returned reason-code set contains its targeted diagnostic code. A2A-CaseVerify adds offline case-level evidence-support verification to A2A interoperability.
\end{abstract}


\section{Introduction}
Large language model (LLM) agents increasingly coordinate tasks across applications, services, and organizational boundaries. Agent-to-Agent (A2A) is an interoperability protocol that supports agent discovery and the exchange of tasks, messages, and artifacts~\cite{noauthor_agent2agent_nodate}. These exchanges do not establish whether the evidence supports a workflow claim.

Cross-organization LLM-agent workflows may distribute evidence across agents and services that are unavailable during a later dispute or audit. Because a valid message does not establish a complete, case-consistent, and ordered support path, an offline verifier must reconstruct preserved evidence without live system access. Healthcare is one example: prior work protects medical-file content and privacy through cryptographic encapsulation, anonymization, and controlled access ~\cite{alshammari_efficient_2025,alshammari_ai-driven_2025}. Medical LLM agents increasingly plan, invoke tools, access electronic health records (EHRs), and execute multi-step tasks. The security problem extends beyond file protection to workflow-level evidence of roles, handoffs, identity, authorization, and auditability~\cite{wang_survey_2025,xu_comprehensive_2026,shi_ehragent_2024,jiang_medagentbench_2025,yao_llm-based_2026,booth_accelerating_2026}.

This paper uses a healthcare claim-support workflow as a running example, although the same verification approach can be configured for other cross-organization business processes. In this example, a human doctor initiates a case, DoctorAgent creates an order, LabAgent produces a result, and InsuranceAgent receives a claim-support artifact. The healthcare verifier profile requires the human-initiation anchor. Receipt evidence is required only under the receipt-required profile configuration. A well-formed final message can still be unsupported if a required event is absent, belongs to another case, or violates the dependency order.

We consider retrospective verification after a suspicious workflow outcome, dispute, or audit trigger. Preserved evidence consists of relevant A2A objects, business events, and dependency links retained in a portable bundle with recomputable commitments. The verifier may lack live access to the agents, A2A servers, or original logs. We therefore ask: \emph{How can preserved cross-organization A2A workflow evidence be cryptographically committed and later reconstructed offline to determine whether the selected verifier profile is satisfied?}

A2A-CaseVerify treats the preserved bundle as a single case-level verification object with recomputable commitments. It projects preserved runtime objects and business events into a directed case evidence graph. It commits A2A envelopes, structured event records, and predecessor/dependency links through deterministic canonicalization and hash-based commitments. It then combines event and edge leaves into a deterministic case Merkle root. The offline verifier reconstructs the graph and returns \textsc{SUPPORTED}/\texttt{OK} only when event integrity, dependency-link integrity, case membership, support-linkage, ordering consistency, and profile-required evidence all hold.

\noindent\textbf{Research gap and novelty.} Existing A2A, receipt, and provenance mechanisms support interoperable exchanges and verifiable records~\cite{noauthor_agent2agent_nodate,alshammari_authenticated_2026,noauthor_peac_2026,world_wide_web_consortium_w3c_prov-dm_nodate}. The remaining gap addressed in this paper is a case-level decision that jointly checks required events, case membership, predecessor links, and order for a final workflow claim. A2A-CaseVerify addresses this gap by combining cryptographic commitments with profile-based offline reconstruction of the case evidence path.

\noindent\textbf{Contributions.} This paper makes four contributions:
\begin{itemize}
\item \textbf{Case evidence graph:} We define a directed case evidence graph for a disputed cross-organization A2A workflow. 
\item \textbf{Merkle-linked evidence binding:} We commit A2A envelopes, event records, and predecessor edges; event and edge leaves form a deterministic case Merkle root. 
\item \textbf{Reason-coded offline verification:} We define verifier semantics that distinguish commitment mismatches (protection failures) from intact evidence that does not satisfy the selected profile (reconstruction failures).
\item \textbf{Prototype validation:} We evaluate one canonical supported healthcare-profile bundle and 13 controlled mutations that target protection and reconstruction checks. We also test runtime-object normalization using a separate valid bundle generated with the official Python A2A SDK.
\end{itemize}

\section{Related Work}
A2A~\cite{noauthor_agent2agent_nodate} provides the runtime
interaction layer. However, the specification does not define an offline case-level evidence-support verifier. Existing work primarily addresses deployment-oriented A2A security, including Agent Card management, task-execution integrity, and authentication, or compares A2A with other agent interoperability protocols~\cite{narajala_building_2025,ehtesham_survey_2025}. At the evidence layer, prior message-level A2A work verifies individual messages and log inclusion using canonical representation, payload commitments, Merkle proofs, and signed checkpoints~\cite{alshammari_authenticated_2026}. A2A-CaseVerify instead evaluates preserved evidence after ingestion to determine whether a final workflow claim is supported by the required case-level evidence path.

Portable receipt systems such as PEAC provide signed, offline-verifiable interaction records and A2A metadata mappings~\cite{noauthor_peac_2026,noauthor_peac_2026-1}. Merkle transparency systems such as Certificate Transparency, Rekor/Sigstore, and SCITT/RFC~9943 provide append-only registration, inclusion proofs, signed statements, transparency services, and receipts~\cite{laurie_certificate_2021,sigstore_rekor_nodate,h_birkholz_architecture_2026}. In-toto verifies signed supply-chain layouts and authorized functionary steps~\cite{torres-arias_-toto_2019}. A2A-CaseVerify evaluates a graph of preserved A2A runtime objects and business events against a case-specific evidence-dependency profile offline. 

W3C PROV and PROV-AGENT model provenance entities, activities, agents, and agentic workflow interactions~\cite{world_wide_web_consortium_w3c_prov-dm_nodate,souza_prov-agent_2025}. Workflow-provenance research shows that provenance can support transparency, reproducibility, and accountability in trustworthy AI workflows~\cite{souza_workflow_2024}. AgentSec provides a diagnostic dataset that records agent decisions, tool interactions, memory updates, and provenance relations~\cite{hmimou_dataset_2026}. ProvenanceTrace is a vision and proof-of-concept study that decomposes generated outputs into atomic claims and links each claim to supporting evidence~\cite{saxena_provenancetrace_2026}. These works enrich agentic audit trails but do not verify whether a preserved A2A case graph is dependency-complete under a case-specific profile. Conformance checking compares observed executions with expected process models~\cite{dunzer_conformance_2019}. A2A-CaseVerify jointly checks cryptographic bindings and profile-defined evidence dependencies in a preserved A2A case graph.

\section{Problem, Assumptions, and Failure Model}
The formal verifier receives four inputs: a preserved evidence bundle, a target case identifier $c$, an expected case root $R_c$, and a trusted verifier profile $P$. The expected root is a trusted Merkle commitment recorded when the evidence bundle was preserved. We define the verifier profile as
\[
P=(C_P,T_P,E_P,Q_P,O_P,G_P,A_P),
\]
where $C_P$ is the case-membership predicate evaluated against $c$, $T_P$ contains the required node types, $E_P$ contains configured evidence requirements, $Q_P$ contains the required typed support-edge patterns, $O_P$ is the workflow-order predicate, $G_P$ is the graph-validity predicate, and $A_P$ identifies the node types that require A2A-envelope binding. Table~\ref{tab:profile} instantiates these components for the healthcare claim-support workflow. An A2A envelope is the canonical set of protocol objects preserved for one runtime exchange. An edge is a typed predecessor/support relation between two event nodes.

The verifier first checks the A2A-envelope, event, edge, and root commitments. It then evaluates case membership, support linkage, ordering, dependency completeness, and required evidence against $P$. The expected root is assumed to come from an authenticated checkpoint or trusted evidence manifest. Recorded envelope hashes also require trusted preservation. They are checked separately and are not authenticated by $R_c$ alone. If the root is attacker-controlled, the verifier cannot provide a meaningful support decision.

The prototype implements $P$ as fixed healthcare checks in the verifier code and assumes that this code is uncompromised. It has no separate semantic profile identifier or version. The commitment-format tag identifies the hash construction, not the semantic profile. Neither the full profile nor its semantic version is bound to $R_c$, and no profile-version check is performed. Replay requires the same rules, since different rules may produce a different decision for unchanged evidence and root.

\begin{figure*}[t]
\centering
\includegraphics[width=0.73\textwidth]
{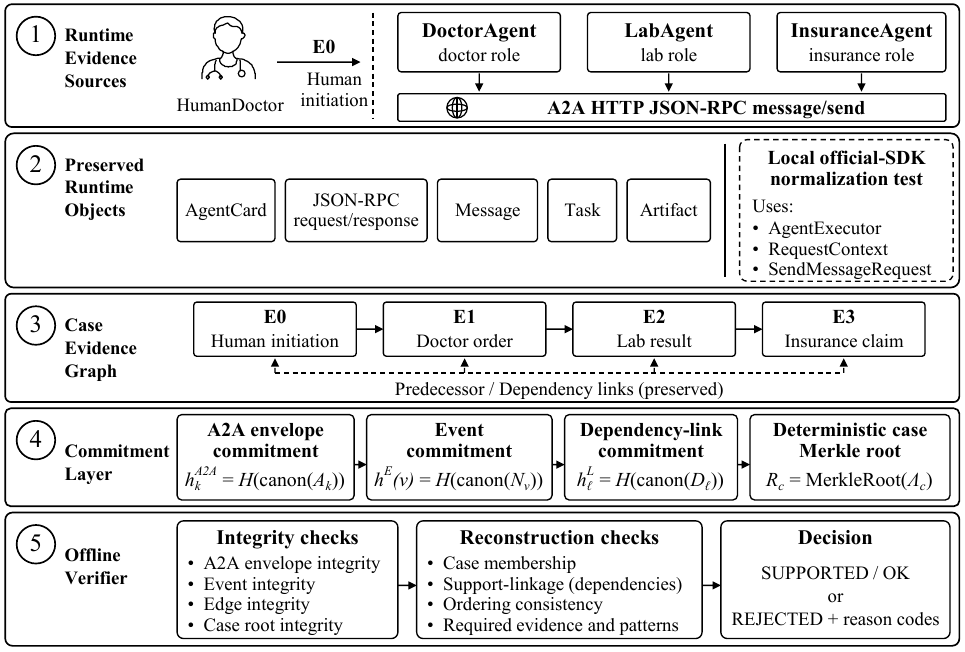}

\caption{A2A-CaseVerify architecture from preserved runtime objects and business events to offline case-graph verification.}
\label{fig:architecture}
\end{figure*}

Fig.~\ref{fig:architecture} separates runtime capture and evidence preservation from the commitment and offline-verification stages.

The adversary may submit an incomplete, reordered, cross-case, or tampered evidence bundle. Possible changes include modifying events, omitting required predecessor evidence, substituting evidence from another case, or altering dependency links. The verifier determines whether the submitted bundle satisfies the selected profile; it does not attribute the cause of a failure.

The prototype uses SHA-256, with digests represented as 64 lowercase hexadecimal characters. Here, $\mathrm{canon}(x)$ is the UTF-8 encoding of Python \texttt{json.dumps(x)} with \texttt{sort\_keys=True}, \texttt{separators=(",",":")}, and \texttt{ensure\_ascii=False}. This deterministic serializer does not claim full RFC~8785 conformance~ \cite{rundgren_json_2020}. The constant $\tau$ denotes the commitment-format tag \path{merkle_linked_case_graph_v1}.

For the healthcare profile, each business-relevant action is represented by an event node. A node type identifies the event category, such as human initiation, doctor order, lab result, or claim submission. A support link is a directed edge $u\rightarrow v$ stating that event $u$ is required evidence for event $v$ under the profile. It represents a business dependency, not merely data transfer or temporal adjacency. For example, doctor order $\rightarrow$ lab result means that the lab-result event depends on the preserved doctor-order event as its required predecessor evidence.

Human initiation is a design choice of the healthcare profile used in this study, not a general A2A or healthcare-workflow requirement. Discussions of human oversight and clinician review in~\cite{european_union_regulation_2024, us_food_and_drug_administration_clinical_2026, toro-tobon_clinician_2026} motivate this profile choice.

The baseline path is $v_H\rightarrow v_D\rightarrow v_L\rightarrow v_I$, where $v_H$, $v_D$, $v_L$, and $v_I$ denote the human-initiation, doctor-order, lab-result, and insurance-claim nodes, respectively. The human-initiation record is not an A2A runtime exchange and does not imply manual approval of every later agent action. It is preserved as a human-originated business event, hashed, linked to the DoctorAgent order by a support edge, and included in the case Merkle root. A bundle that contains downstream A2A evidence but lacks this event is rejected with \texttt{MISSING\_HUMAN\_INITIATION}.

The profile specifies the required node types in $T_P$, such as doctor order and lab result, and the required source-to-target edge patterns in $Q_P$, such as doctor order $\rightarrow$ lab result. Three reconstruction failures are distinguished. \emph{Missing evidence} means that a required node type is absent, such as a missing laboratory result. \emph{Broken dependency} means that the required endpoint events exist, but no committed edge satisfies the required source-to-target pattern. For example, the lab result is not linked to the doctor order. \emph{Wrong order} means that the required events and dependency edge exist, but their committed sequence values violate the profile's ordering constraint. For example, the claim sequence is not later than the lab-result sequence. These reconstruction failures can occur even when all cryptographic protection checks pass.

\begin{table}[t]
\caption{Healthcare claim-support verifier profile}
\label{tab:profile}
\centering
\scriptsize
\begin{tabularx}{\columnwidth}{L{0.36\columnwidth}Y}
\toprule
Profile element & Required condition \\
\midrule

Case membership $C_P$
&
Every node stores the target case identifier $c$.
Every support edge connects endpoint nodes that also
store $c$.
\\

Required node types $T_P$
&
Human-doctor initiation, doctor order, lab result, and
insurance claim submission.
\\

Configured evidence $E_P$
&
Receipt evidence is required only when the receipt-required
profile configuration is selected.
\\

Required edge patterns $Q_P$
&
Human initiation $\rightarrow$ doctor order, doctor order
$\rightarrow$ lab result, and lab result $\rightarrow$
insurance claim submission.
\\

Order rule $O_P$
&
Each node stores a committed logical workflow position
$\mathrm{seq}(v)$. For every required edge $(u,v,\rho)$,
$\mathrm{seq}(u)<\mathrm{seq}(v)$.
\\

Graph validity $G_P$
&
Event identifiers are unique, predecessor endpoints resolve,
and every node is the final claim or one of its direct
or indirect predecessors.
\\

A2A-envelope binding $A_P$
&
DoctorAgent, LabAgent, and InsuranceAgent event nodes
require A2A-envelope binding. The human-initiation node does
not require A2A-envelope binding.
\\

\bottomrule
\end{tabularx}
\end{table}

\section{Merkle-Linked Case Evidence Graph}
For case $c$, each preserved business-relevant event is represented by a node. Each node $v\in V_c$ stores a committed event identifier $\mathrm{id}(v)$, case identifier $c$, node type, and committed logical workflow position $\mathrm{seq}(v)$. A2A-derived events also store runtime-linkage fields. A directed edge $\ell=(u,v,\rho)$ represents a typed predecessor/support relation from source event $u$ to target event $v$, where $\rho$ is the relation-type label.

We define the directed case evidence graph as
\begin{equation}
G_c=(V_c,L_c,\lambda_V,\lambda_L),\quad
L_c\subseteq V_c\times V_c\times\mathcal{R}.
\end{equation}
Here, $V_c$ is the set of event nodes for case $c$, $L_c$ is the set of directed support edges, and $\mathcal{R}$ is the set of relation-type labels. The functions $\lambda_V$ and $\lambda_L$ assign attributes to nodes and edges, respectively. The subset $V_c^{A2A}\subseteq V_c$ contains nodes derived from A2A runtime exchanges. The components $T_P$ and $Q_P$ contain the required node types and typed support-edge patterns, respectively. For example, doctor order and lab result are node types, while doctor order $\rightarrow$ lab result is an edge pattern.

The healthcare case-validation suite assumes one final claim, at most one node per required type, and one intended predecessor path without duplicate edges. Duplicate event identifiers yield \texttt{DUPLICATE\_EVENT\_ID}, and unresolved predecessors yield \texttt{BROKEN\_PREDECESSOR}. When a claim is present, nodes outside its predecessor closure yield \texttt{ORPHANED\_EVIDENCE}. This closure
contains the claim and all events reached by following its predecessor references. The evaluation does not cover multiple same-type candidates or competing paths. The prototype has no dedicated ambiguity-rejection check. Unexpected node types and additional or duplicate edges are not rejected solely for being extra. They remain subject to the other verification checks.

\begin{figure*}[t]
\centering
\includegraphics[width=0.73\textwidth]
{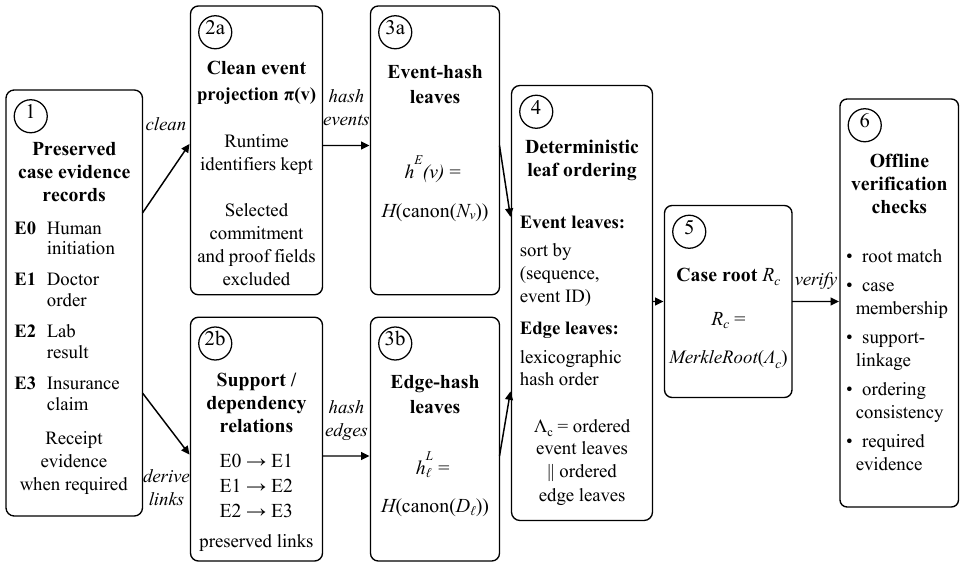}
\caption{Merkle-linked reconstruction from preserved events and dependency links to offline case-profile checks.}
\label{fig:casegraph}
\end{figure*}

Fig.~\ref{fig:casegraph} gives a left-to-right overview of case reconstruction. Preserved events become event-hash leaves, support relations become edge-hash leaves, and both leaf sets form the case root. The reconstructed graph is then checked for case membership, support linkage, event order, and required evidence.

For exchange $x_k$, $\mathrm{Env}_k$ contains the following
JSON fields:
\par
\begingroup
\raggedright
\noindent
\texttt{agent}, \texttt{agent\_card}, \texttt{agent\_card\_sha256}, \texttt{a2a\_rpc\_request}, \texttt{a2a\_request\_message}, \texttt{jsonrpc\_response}, \texttt{a2a\_task\_response}, \texttt{ollama\_model}, \texttt{prompt\_sha256}, and \texttt{derived\_evidence\_event\_id}.
\par
\endgroup
\noindent
Values are copied from the preserved trace; absent fields
are represented by JSON \texttt{null}. The hash input $A_k$
contains exactly three fields: \texttt{profile} set to $\tau$,
\texttt{kind} set to \texttt{a2a\_envelope}, and
\texttt{envelope} set to $\mathrm{Env}_k$.
The envelope commitment is
\begin{equation}
h^{A2A}_k=H(\mathrm{canon}(A_k)).
\end{equation}

The A2A-derived event stores this envelope hash and the related runtime identifiers. This association is called the A2A-envelope binding.

For event $v$, $\pi(v)$ retains all top-level event fields
except \texttt{a2a\_envelope\_hash}, \texttt{event\_hash},
\texttt{predecessor\_edge\_hashes},
\texttt{case\_merkle\_root},
\texttt{case\_root\_leaf\_count},
\texttt{protection\_profile},
\texttt{merkle\_leaf\_index}, and \texttt{merkle\_proof}.
Case and event identifiers, predecessor references, payload
commitments, and runtime-linkage fields therefore remain.
Let $N_v$ contain exactly \texttt{profile}:$\tau$,
\texttt{kind}:\texttt{event\_node}, and
\texttt{event}:$\pi(v)$.
\begin{equation}
h^E(v)=H(\mathrm{canon}(N_v)).
\end{equation}

The \texttt{payload\_sha256} field hashes the first
Artifact's first data Part for controlled agent events,
and the retained \texttt{payload} object for SDK events.
Both helpers use SHA-256 over UTF-8, key-sorted compact
JSON; the controlled helper uses \texttt{ensure\_ascii=True}
and the SDK helper uses \texttt{False}.
For controlled human initiation, the hashed object copies
\texttt{case\_id}, \texttt{actor}, \texttt{actor\_type},
\texttt{actor\_role}, and \texttt{event\_type} from the
event, with \texttt{action}=\texttt{initiate\_case},
\texttt{interface}=\texttt{doctor\_web\_interface}, and
\texttt{profile\_requirement}=
\path{human_in_the_loop_healthcare_claim_profile}.
The SDK human event retains its hashed object in
\texttt{payload}.

The prototype uses relation type \texttt{predecessor}
for edge $\ell=(u,v,\rho)$. Its JSON object $D_\ell$
contains exactly five fields:
\par
\begingroup
\raggedright
\noindent
\texttt{profile} set to $\tau$,
\texttt{kind} set to \texttt{predecessor\_edge},\\
\texttt{relation\_type} set to $\rho$,\\
\texttt{predecessor\_event\_hash} set to $h^E(u)$,\\
\texttt{current\_event\_hash} set to $h^E(v)$.
\par
\endgroup
\noindent
The edge commitment is
\begin{equation}
h^L_\ell=H(\mathrm{canon}(D_\ell)).
\end{equation}
Case and event identifiers are covered through the endpoint
hashes. No matched profile rule is included in $D_\ell$.
For an unresolved predecessor, preservation instead hashes
a JSON object with exactly four fields:
\par
\begingroup
\raggedright
\noindent
\texttt{profile} set to $\tau$,\\ \texttt{kind} set to \texttt{broken\_predecessor\_edge},\\ \texttt{missing\_predecessor\_id} set to the unresolved identifier,\\ \texttt{current\_event\_hash} set to $h^E(v)$.
\par
\endgroup
\noindent 
Verification rejects the unresolved link.

A support link states that source event $u$ is required evidence for target event $v$. It is not merely an information-flow arrow, temporal adjacency, or A2A conversation reference. The committed value $\mathrm{seq}(v)$ represents logical workflow position; it is not an event timestamp or a Merkle-leaf position.  An edge $\ell=(u,v,\rho)$ satisfies a required profile pattern only if its endpoint events exist, both use case identifier $c$, its types match that pattern, and $\mathrm{seq}(u)<\mathrm{seq}(v)$. For example, the doctor-order event must have a smaller sequence value than the lab-result event.

A2A correlation identifiers bind runtime evidence to an event but do not by themselves establish business-level support. Event hashes protect event contents, edge hashes protect the asserted dependency relations, and profile $P$ determines whether those relations support the final claim. Individually valid events may therefore still fail when a required predecessor relation is missing, changed, or not allowed by the profile.

The deterministic case Merkle root contains one event-hash leaf for each preserved event and one dependency-link leaf for each preserved edge. Event leaves are ordered by increasing workflow-sequence value, with the event identifier used to break ties. Edge leaves are ordered lexicographically by their hash values:
\begin{equation}
\Lambda_c=
[h^E(v):v\in \mathrm{sort}_{(\mathrm{seq},\mathrm{id})}(V_c)]
\,\Vert\,
\mathrm{sort}_{\mathrm{lex}}
([h^L_\ell:\ell\in L_c]).
\end{equation}
The case root is
\begin{equation}
R_c=\mathrm{MerkleRoot}(\Lambda_c).
\end{equation}

A Merkle parent is $H(\mathrm{canon}(M))$, where $M$
contains exactly \texttt{profile}:$\tau$,
\texttt{kind}:\texttt{merkle\_parent},
\texttt{left}:$a$, and \texttt{right}:$b$ for hexadecimal
child strings $a,b$. Thus, children are JSON string fields,
not concatenated raw bytes. Adjacent leaves are paired
from left to right, duplicating the last leaf at odd-sized
levels; a single leaf is returned unchanged. An empty tree
hashes the object containing only \texttt{profile}:$\tau$
and \texttt{kind}:\texttt{empty\_case\_root}.
Verification checks recorded event and edge hashes
separately before rebuilding the root from those recorded
leaves.

This deterministic leaf ordering is used only to reproduce the same Merkle root. It does not establish valid workflow order. After reconstructing the graph, the verifier applies the sequence rule above to each profile-required support edge and returns \texttt{WRONG\_ORDER} when the rule is violated.

 The verifier records each detected failure in a diagnostic reason-code set $R$. It checks every event commitment and case identifier, every A2A-derived event’s envelope binding, and every edge’s commitment and endpoints. It recomputes the case root once from the complete event-and-edge leaf set, then applies the profile’s required-node, support-link, and sequence rules.

\begin{equation}
\operatorname{Decision}_P(G_c)=
\begin{cases}
\textsc{SUPPORTED}, & R=\emptyset,\\
\textsc{REJECTED}, & R\neq\emptyset.
\end{cases}
\end{equation}

When $R=\emptyset$, the verifier returns
\textsc{SUPPORTED} with status \texttt{OK}. When
$R\neq\emptyset$, it returns \textsc{REJECTED} with the diagnostic reason-code set $R$. Protection failures map to A2A-envelope, event-commitment, edge-commitment, or case-root mismatch codes, while reconstruction failures map
to case-mismatch, broken-predecessor, wrong-order, or missing-evidence codes. In the healthcare profile, absence of the required human-doctor initiation node maps specifically to
\texttt{MISSING\_HUMAN\_INITIATION}.

\begin{algorithm}[t]
\caption{Offline Secure Reconstruction and Verification}
\label{alg:verify}
\begin{algorithmic}[1]
\Require Preserved bundle, case $c$, expected root $R_c$, trusted profile $P$
\Ensure \textsc{SUPPORTED}/\texttt{OK} or \textsc{REJECTED} with reason codes

\State Initialize reason-code set $R \gets \emptyset$
\State Parse A2A exchanges, evidence nodes, and predecessor edges
\State Recompute A2A-envelope, event, edge, and root commitments
\State Let $R_c^{\mathrm{rec}}$ be the recomputed case root

\If{any A2A envelope mismatch}
    \State Add \texttt{A2A\_ENVELOPE\_MISMATCH} to $R$
\EndIf

\If{any event commitment mismatch}
    \State Add \texttt{EVENT\_COMMITMENT\_MISMATCH} to $R$
\EndIf

\If{any edge commitment mismatch}
    \State Add \texttt{EDGE\_COMMITMENT\_MISMATCH} to $R$
\EndIf

\If{$R_c^{\mathrm{rec}} \neq R_c$}
    \State Add \texttt{CASE\_ROOT\_MISMATCH} to $R$
\EndIf

\State Reconstruct $G_c=(V_c,L_c,\lambda_V,\lambda_L)$

\If{case membership fails}
    \State Add \texttt{CASE\_MISMATCH} to $R$
\EndIf

\If{the parsed events contain duplicate event identifiers}
    \State Add \texttt{DUPLICATE\_EVENT\_ID} to $R$
\EndIf

\If{required support linkage fails}
    \State Add \texttt{BROKEN\_PREDECESSOR} to $R$
\EndIf

\If{ordering consistency fails}
    \State Add \texttt{WRONG\_ORDER} to $R$
\EndIf

\If{a claim exists and some event is outside its predecessor closure}
    \State Add \texttt{ORPHANED\_EVIDENCE} to $R$
\EndIf

\If{profile-required human initiation is missing}
    \State Add \texttt{MISSING\_HUMAN\_INITIATION} to $R$
\EndIf

\If{any profile-required node type other than human initiation, or any configured receipt evidence, is missing}
    \State Add the corresponding missing-evidence reason code to $R$
\EndIf

\If{$R=\emptyset$}
    \State \Return \textsc{SUPPORTED}/\texttt{OK}
\Else
    \State \Return \textsc{REJECTED} with reason codes $R$
\EndIf
\end{algorithmic}
\end{algorithm}
\section{Analysis}
A2A-CaseVerify uses two layers. The protection layer recomputes A2A-envelope, event, edge, and root commitments. The reconstruction layer checks whether the protected evidence supports the claimed path under profile $P$. If a committed event, predecessor relation, or Merkle leaf set is modified after preservation, the corresponding recomputed commitment changes except under a hash collision. The reconstruction layer rejects a bundle if a record belongs to another case, a required predecessor edge is absent, or a sequence violates the profile. This applies even when individual event hashes are valid. A \textsc{SUPPORTED} result means evidentiary support under profile $P$.

Event integrity alone does not establish profile satisfaction. A valid doctor order and valid lab result may still fail if the lab result was produced for another order or no predecessor relation is preserved. Event hashes protect the preserved predecessor references. Edge commitments additionally bind the relation type and endpoint event hashes, which cover their case and event identifiers. Profile $P$ is applied during reconstruction. Its semantic rules are not inputs to the implemented edge commitment.

Let $q$ be the number of distinct canonical inputs hashed during verification and $m$ the hash length. Under the birthday-bound approximation, the accidental collision probability is approximately $q(q-1)/2^{m+1}$. For 256-bit hashes and the controlled bundles used here, this probability is negligible. Deterministic canonicalization is equally necessary because the verifier must reproduce the same hash inputs without live access to organizational systems.

\section{Prototype and Evaluation}

\subsection{Validation Strategy}
Validation checks each preserved bundle against three references: the recorded envelope, event, and edge commitments; the expected case root; and the selected verifier profile $P$. The commitments detect modified evidence. The expected root detects changes to the committed leaf set. The profile defines the required events, support links, case membership, and workflow order. The 14-scenario controlled suite contains one canonical supported healthcare-profile bundle and 13 controlled mutations. Each case has an expected decision, and each mutation has a targeted diagnostic code. 

For the controlled workflow, the prototype checks the runtime-evidence boundary against the evidence schema defined in this paper. It verifies that the preserved bundle contains AgentCard data, JSON-RPC \texttt{message/send} request/response objects, Message, Task, and Artifact. It also checks for runtime metadata and identifiers linking A2A-derived events to their runtime objects. Validation also includes a separate local official-SDK workflow to test normalization into the evidence-event schema.

\begin{figure*}[t]
\centering
 \includegraphics[width=0.67\textwidth]
{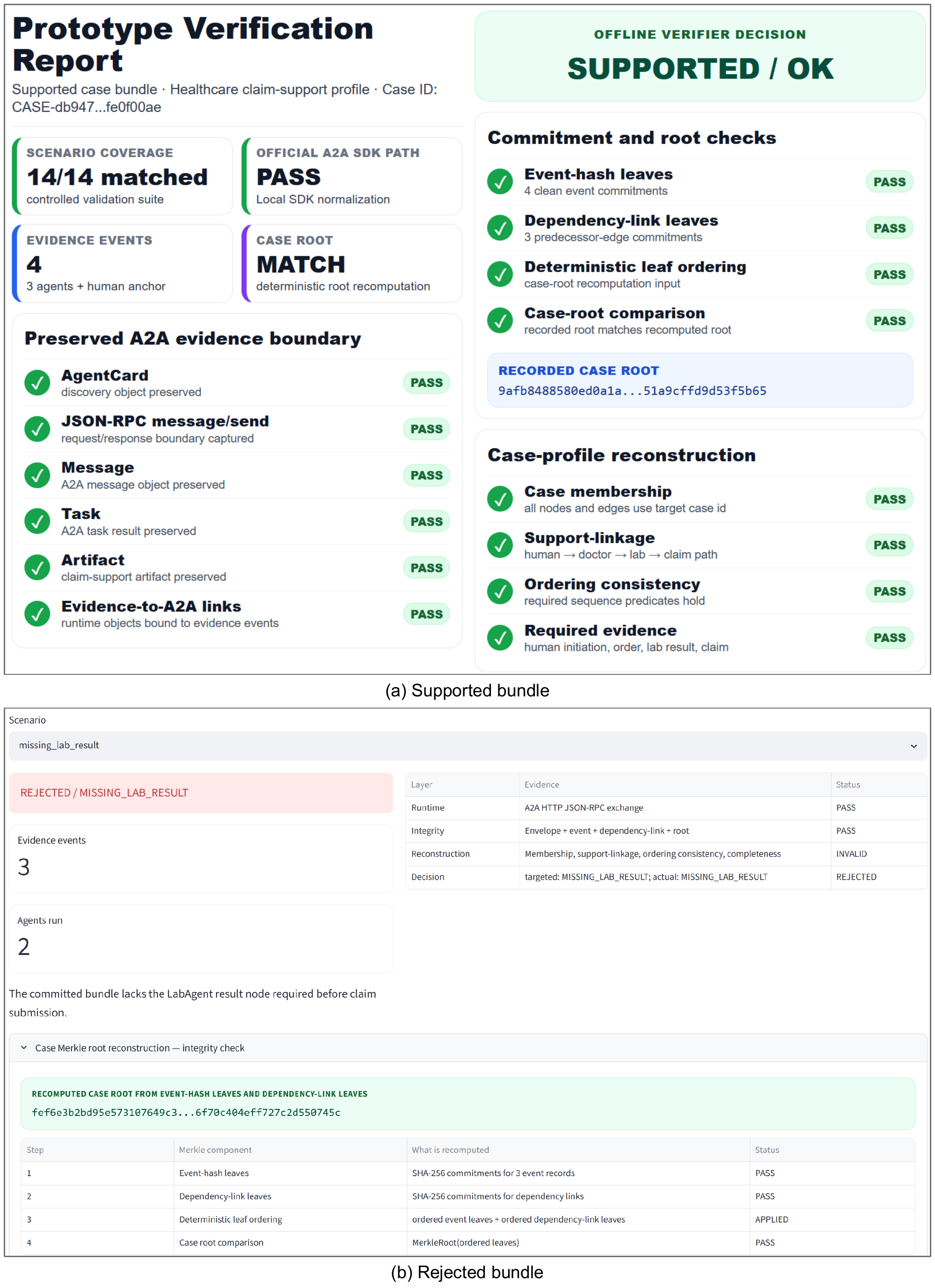}
\caption{Reports for (a) a supported bundle and (b) a \texttt{missing\_lab\_result} bundle that passes integrity and root checks but fails profile reconstruction.}

\label{fig:dashboard}
\end{figure*}

Fig.~\ref{fig:dashboard} shows reports for a supported bundle (a) and a \texttt{missing\_lab\_result} reconstruction failure (b).

\subsection{Implementation Details}
The implementation separates runtime capture from offline verification. Runtime capture records A2A HTTP JSON-RPC exchanges and links them to evidence events. The prototype verifies the preserved bundle and runtime trace against the expected root using its fixed healthcare checks. A2A-originated events retain identifiers for message, task, artifact, context, and request/response objects. Human initiation is preserved with the same case identifier and a sequence preceding the DoctorAgent order, then linked by a predecessor edge.

The verifier recomputes commitments from clean event projections. It uses the exact projection $\pi(v)$ defined in Section~IV. Reconstruction runs after commitment recomputation because a cryptographically intact bundle may still be structurally insufficient. The verifier returns all reason codes to support diagnosis of compound failures.

The controlled workflow uses HTTP JSON-RPC with DoctorAgent, LabAgent, and InsuranceAgent. This path uses a custom \texttt{message/send} implementation. The valid case contains four protected evidence events over three agent roles and one human-initiation anchor. The separate local test used Python \texttt{a2a-sdk} 1.1.0 and the A2A Specification 1.0 data model~ \cite{noauthor_agent2agent_nodate}. It invoked \texttt{AgentExecutor.execute} through \texttt{RequestContext}, not through an HTTP server. Per exchange, one \texttt{SendMessageRequest} supplied one user \texttt{Message}. The returned \texttt{Task} contained one \texttt{Artifact} with one text \texttt{Part}. The SDK request message is copied to \path{a2a_rpc_request.params.message}. The normalized Task is copied to \path{a2a_task_response}, and AgentCard data is copied to \path{agent_card}. The normalized Task contains the input message in \texttt{history}, plus the returned Artifact and text Part. The message, task, artifact, and context identifiers populate \path{a2a_message_id}, \path{a2a_task_id}, \path{a2a_artifact_id}, and \path{a2a_context_id} in the event, respectively. The separate \path{a2a_request_message} and \path{jsonrpc_response} fields are absent from this SDK trace and become JSON \texttt{null} in $\mathrm{Env}_k$. The JSON-RPC request wrapper was constructed for normalization, not captured from an SDK network exchange. \texttt{RequestContext} and \texttt{AgentExecutor} are execution components, not separately serialized envelope objects. The normalized events are processed by the same Merkle-linked verifier. Other response shapes were not evaluated.

For reconstruction scenarios, the incomplete or profile-invalid bundle is created first, then its commitments and expected root are computed and recorded. Thus, the \texttt{missing\_lab\_result} bundle can match its recorded root while failing the profile. Protection scenarios instead modify evidence after the relevant commitments and expected root are fixed. Unresolved predecessors may also trigger an edge-commitment diagnostic, so protection and reconstruction diagnostics can overlap.

\begin{table*}[!t]
\caption{Controlled validation scenarios and targeted diagnostic codes}
\label{tab:scenario_validation}
\centering
\scriptsize
\renewcommand{\arraystretch}{0.96}
\setlength{\tabcolsep}{3pt}
\begin{tabularx}{\textwidth}{@{}L{0.20\textwidth} L{0.37\textwidth} L{0.18\textwidth} Y@{}}
\toprule
Scenario & Controlled condition & Targeted check & Targeted code \\
\midrule

Valid claim &
Complete path with required nodes and links; case root matches &
All checks &
\texttt{OK} \\

Missing human initiation &
Human-initiation anchor absent; downstream evidence remains &
Human-initiation requirement &
\texttt{MISSING\_HUMAN\_INITIATION} \\

Missing doctor order &
DoctorAgent order absent &
Required node &
\texttt{MISSING\_DOCTOR\_ORDER} \\

Missing lab result &
LabAgent result absent &
Required node &
\texttt{MISSING\_LAB\_RESULT} \\

Missing required receipt &
Configured receipt evidence absent &
Configured evidence &
\texttt{MISSING\_RECEIPT} \\

Post-commitment case-ID change &
Case-ID field changes after event commitment &
Event commitment &
\texttt{EVENT\_COMMITMENT\_MISMATCH} \\

Cross-case committed event &
Committed event from another case is included &
Case membership &
\texttt{CASE\_MISMATCH} \\

Wrong workflow order &
Required nodes and links remain; sequence violates profile order &
Workflow order &
\texttt{WRONG\_ORDER} \\

Broken predecessor link &
Endpoint events remain; predecessor reference is unresolved &
Predecessor linkage &
\texttt{BROKEN\_PREDECESSOR} \\

Payload tampering &
Payload-hash field changes after event commitment &
Event commitment &
\texttt{EVENT\_COMMITMENT\_MISMATCH} \\

Dependency-link tampering &
Dependency-link endpoint changes after edge commitment &
Edge commitment &
\texttt{EDGE\_COMMITMENT\_MISMATCH} \\

Event removed after root &
Lab-result event is removed after recording the expected root &
Case-root equality &
\texttt{CASE\_ROOT\_MISMATCH} \\

Sequence-field tampering &
Claim sequence changes after event commitment &
Event commitment &
\texttt{EVENT\_COMMITMENT\_MISMATCH} \\

A2A-envelope tampering &
Envelope component changes after envelope commitment &
Envelope commitment &
\texttt{A2A\_ENVELOPE\_MISMATCH} \\

\bottomrule
\end{tabularx}
\renewcommand{\arraystretch}{1.0}
\end{table*}

All 14 scenarios matched their expected decisions. For each rejected case, the returned set $R$ contained its targeted diagnostic code. Table~\ref{tab:scenario_validation} lists test targets, not selected primary codes. The verifier returns all applicable codes. For example, removing an event after recording the case root yields \texttt{CASE\_ROOT\_MISMATCH} and may also expose missing evidence or broken links.

\subsection{Discussion of Validation Results}
Reconstruction predicates rejected the missing-evidence, case-mismatch, broken-predecessor, and wrong-order cases. Protection predicates rejected the envelope-, event-, dependency-link-, and root-tampering cases. The two layers provide complementary checks.

Table~\ref{tab:verifier_capability} presents a logical failure-coverage analysis across four verifier scopes. Message-only checks commitments to individual preserved A2A exchanges. Event-root checks event commitments and their root. The Event+edge scope additionally checks dependency-link commitments. The Event-root and Event+edge scopes exclude A2A-envelope binding and full profile reconstruction. A2A-CaseVerify combines envelope, event, edge, root, and profile checks. Each entry is derived logically from the checks defined for its scope. The missing-evidence and wrong-order rows assume internally consistent commitments and a recomputed root that matches the expected root. Event hashes can expose changes to predecessor references when those references are included in the committed event.

\begin{table}[t]
\caption{Failure classes visible under different verifier scopes}
\label{tab:verifier_capability}
\centering
\scriptsize
\renewcommand{\arraystretch}{0.98}
\setlength{\tabcolsep}{1.6pt}
\begin{tabular*}{\columnwidth}{@{\extracolsep{\fill}}L{0.40\columnwidth}C{0.11\columnwidth}C{0.11\columnwidth}C{0.12\columnwidth}C{0.16\columnwidth}@{}}
\toprule
Failure class &
\makecell{Message\\only} &
\makecell{Event\\root} &
\makecell{Event\\+ edge} &
\makecell{A2A-\\CaseVerify} \\
\midrule
Missing required evidence & \no & \no & \no & \yes \\
Cross-case valid hash & \no & \partly & \partly & \yes \\
Wrong workflow order & \no & \no & \no & \yes \\
Broken predecessor & \no & \no & \partly & \yes \\
Tampered event content & \no & \yes & \yes & \yes \\
Tampered dependency link & \no & \partly & \yes & \yes \\
A2A-envelope binding mismatch & \partly & \no & \no & \yes \\
\bottomrule
\end{tabular*}
\vspace{2pt}
\parbox{\columnwidth}{\scriptsize\emph{Note:}
\yes{} = detectable by the stated checks,
\no{} = not detectable by those checks, and
\partly{} = conditionally detectable, depending on committed fields or configured checks.}
\end{table}

\subsection{Performance Measurements}

Measurements used a Surface Laptop Studio 2 with an Intel
Core i7-13700H, 32~GB RAM, Windows 11, and Python 3.11. The two experiments measure different verification functions. Both exclude evidence generation, network activity, and file output.

\begin{figure}
\includegraphics[width=\columnwidth]
{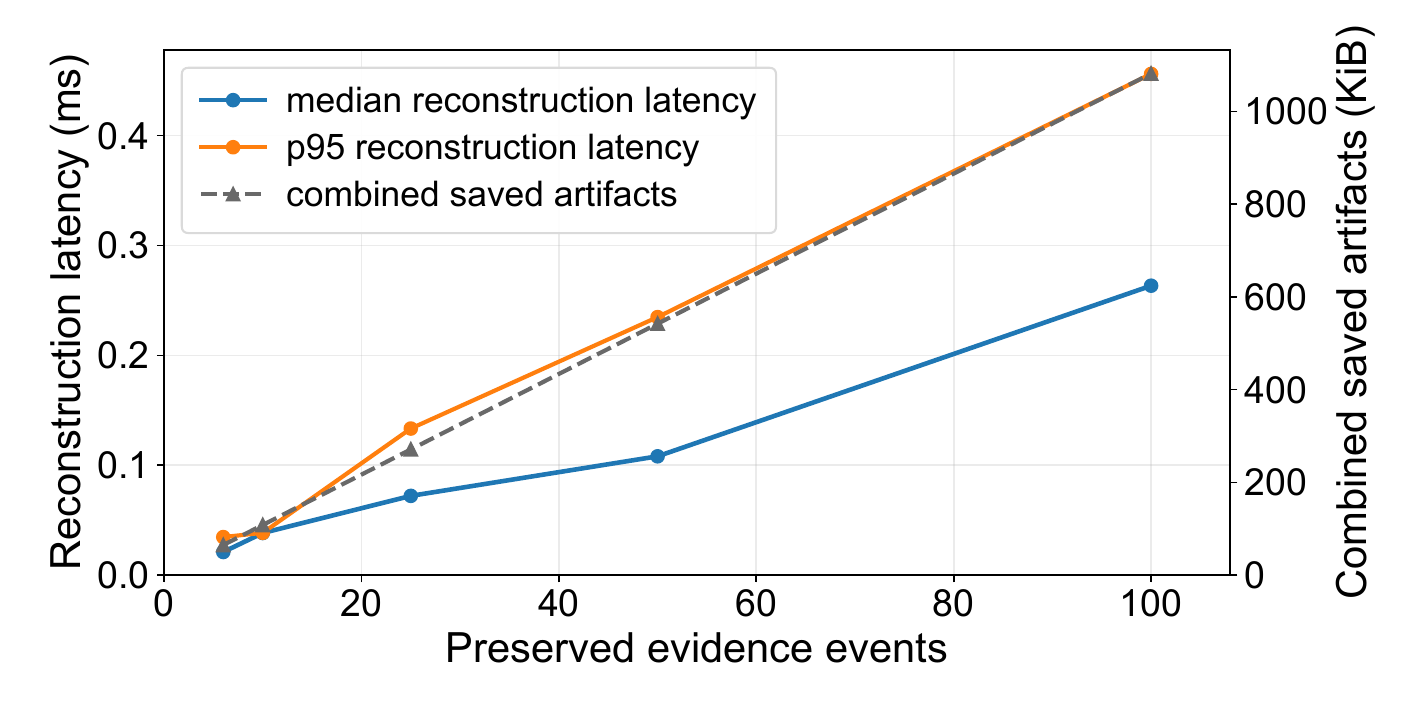}
\caption{Reconstruction-only timing and combined saved-artifact
size for generated chains, with three runs per event count.}
\label{fig:performance_scaling}
\end{figure}

Fig.~\ref{fig:performance_scaling} reports reconstruction-only timings for generated linear chains with $n=6,10,25,50,100$ events and $n-1$ predecessor links. Each size has three runs. Python \texttt{perf\_counter()}
times \texttt{verify\_evidence\_bundle}, covering case membership, predecessor linkage, ordering, and required evidence. Envelope, event, edge, and root recomputation are outside this measured path. At 100 events, the median and p95 are 0.263~ms and 0.456~ms. The median combined size of the saved workflow record, evidence bundle, and verifier report is approximately 1.06~MiB.

Table~\ref{tab:performance_stats} reports the separate \texttt{verify\_claim\_bundle} pipeline, which checks event evidence, checkpoints and witnesses, policy, and claim dependencies. This pipeline is distinct from the Merkle-linked verifier in Algorithm~\ref{alg:verify}.
Python \texttt{perf\_counter\_ns()} times 30 repetitions for each of 15 cases, following one untimed verification per case. The valid claim bundle contains six events. These 450 timed calls are separate from the 14 validation scenarios in Table~\ref{tab:scenario_validation}. The overall
median and p95 are 1.085~ms and 1.791~ms. Saved validation decisions match expectations for all 15 cases.

The reported p95 values are reproduced by linear
interpolation at position $0.95(N-1)$ in the sorted samples. Sample indices start at zero, and $N$ is the number of samples. With three samples per size, Figure~\ref{fig:performance_scaling}
provides descriptive p95 values. These timings apply only to the verification paths measured in each experiment.

\begin{table}[t]
\caption{Claim-verification pipeline timing statistics}
\label{tab:performance_stats}
\centering
\scriptsize
\renewcommand{\arraystretch}{1.02}
\setlength{\tabcolsep}{0pt}

\begin{tabularx}{\columnwidth}{
@{}
L{0.28\columnwidth}
r@{\hspace{5pt}}
r@{\hspace{7pt}}
r@{\hspace{4pt}}
r@{\hspace{4pt}}
r@{\hspace{4pt}}
r@{\hspace{7pt}}
>{\raggedleft\arraybackslash}X
@{}
}
\toprule

Timing group
& Cases
& Runs
& \multicolumn{4}{c}{Latency (ms)}
& \makecell{Size\\(KiB)} \\

\cmidrule(lr){4-7}

&
&
&
Mean
& SD
& Median
& p95
& \\

\midrule

Valid case
& 1
& 30
& 1.545
& 0.294
& 1.496
& 2.059
& 7.17 \\

Core reconstruction
& 7
& 210
& 1.205
& 0.343
& 1.082
& 1.948
& 6.19--7.19 \\

Evidence prerequisites
& 3
& 90
& 0.699
& 0.204
& 0.562
& 1.008
& 6.81--7.18 \\

Profile extensions
& 4
& 120
& 1.197
& 0.125
& 1.212
& 1.454
& 7.18--8.17 \\

\midrule

Overall
& 15
& 450
& 1.124
& 0.354
& 1.085
& 1.791
& 6.19--8.17 \\

\bottomrule
\end{tabularx}

\renewcommand{\arraystretch}{1.0}
\end{table}

\section{Limitations and Considerations}
\label{sec:limitations}

The evaluation uses one healthcare profile, synthetic evidence, and controlled mutations. It excludes deployment traces, independently generated failures, and an external benchmark. A \textsc{SUPPORTED} result establishes profile satisfaction. It does not establish workflow or LLM-output correctness. It also does not independently authenticate agent identity, message origin, or runtime authorization. The decision relies on the trusted-capture, authenticated-root, and trusted-profile assumptions in Section~III.

\section{Conclusion}
\label{sec:conclusion}

This paper presented A2A-CaseVerify, which verifies A2A-envelope commitments separately from the Merkle-linked graph of business events and typed predecessor links. The graph is checked against a trusted case root and the verifier profile. In the tested healthcare cases, the canonical and normalized SDK bundles returned \textsc{SUPPORTED}/\texttt{OK}. All 13 manipulated bundles were rejected, and each returned reason-code set contained its targeted diagnostic code. The results show that message validity and cryptographic integrity alone do not establish final-claim support. Dependency completeness, case membership, ordering, support linkage, and profile-required evidence are also required. Future work will examine larger workflows, randomized failures, profile governance, stronger ingestion trust anchors, and reduced-disclosure verification.

\bibliographystyle{IEEEtran}
\bibliography{Paper3A}
\end{document}